\documentstyle[12pt,righttag]{amsart}

\newtheorem{prop}{Proposition}[section]

\theoremstyle{remark}
\newtheorem{rem}{Remark}[section]

\numberwithin{equation}{section}
\begin{document}

\title[On a class of Dynamical Systems\dots] {On a class of Dynamical Systems
both quasi-bi-Hamiltonian and bi-Hamiltonian} 

\author[C. Morosi]{C. Morosi \dag}
\address {\dag C. Morosi, Dipartimento di Matematica, Politecnico di Milano,
Piazza L. Da Vinci 32, I-20133 Milano, Italy}
\email{carmor@@mate.polimi.it}

\author[G. Tondo]{G. Tondo \ddag}
\address {\ddag G. Tondo, on leave from Dipartimento di Scienze Matematiche,
Universit\`a degli Studi di Trieste, Piaz.le Europa 1, I-34127 Trieste, Italy.}
\email{tondo@@univ.trieste.it}

\date{\dag Dipartimento di Matematica, Politecnico di Milano, Piazza L. Da
Vinci 32,  I-20133 Milano, Italy \\
\ddag Department of Applied Mathematical Studies, University of Leeds, \\
Leeds LS2 9JT, United Kingdom} 

\subjclass {Primary 58F07; Secondary 58F05, 70H20}

\maketitle

\begin{abstract} It is shown that a class of dynamical systems (encompassing
the one recently considered by F. Calogero in \cite{C1}) is both
quasi-bi-Hamiltonian and  bi-Hamiltonian. The first formulation entails the
separability of these systems; the second one is obtained trough a non canonical
map whose form is directly suggested by the associated Nijenhuis tensor.
\end{abstract}

\pagebreak

\section {Introduction and Preliminaries} \label{sec:intro} 

The aim of this Letter is to point out some properties of a class of dynamical
systems which admit both a quasi-bi-Hamiltonian (QBH) formulation and a
bi-Hamiltonian (BH) formulation.
\par Let $M$ be an even dimensional differentiable manifold (dim $M=2n$),
$TM$ and $ T^*M$ its tangent and cotangent bundle: a bi-Hamiltonian structure
on $M$  is a pair $(P_0,P_1)$ ($P_0$, $P_1:T^*M\mapsto TM$) of two
compatible Poisson tensors \cite{Magri1}. If $P_0$ is invertible and the
Nijenhuis	tensor $N:= P_1\, P_0^{-1}$ has $n$ functionally independent
eigenvalues $(\lambda_1, \ldots , \lambda_n)$ one can introduce a set of
canonical coordinates $({\boldsymbol\lambda};{\boldsymbol\mu)}$ ($
{\boldsymbol \lambda}:=(\lambda_1,\ldots, \lambda_n); {\boldsymbol
\mu}:=(\mu_1,\ldots,\mu_n)$) referred to as Darboux-Nijenhuis coordinates
\cite{TMagri} such that $P_0$, $P_1$ and $N$ take the matrix form 

\begin{equation} \label{eq:PND} P_0=
\begin{bmatrix} 0& I\\\ - I&0
\end{bmatrix} \ , \qquad P_1 =
\begin{bmatrix} 0&\Lambda\\\ -\Lambda &0
\end{bmatrix} \ , \qquad N=
\begin{bmatrix}
\Lambda&0 \\\ 0&\Lambda
\end{bmatrix} \ ,
\end{equation} where $ I$ denotes the $n \times n$ identity matrix and $ 
\Lambda:=diag(\lambda_1, \ldots, \lambda_n)$. The above form of the BH
structure will be referred to as a $canonical$ BH structure. \par We recall that
the local model of a BH structure was studied in \cite{Turiel}, where it was
shown that a compatible pair $(P_0, P_1)$ of Poisson tensors (being $P_0$
invertible) admits a representation with $P_0$ constant and $P_1$ depending
linearly on the coordinates; the Darboux--Nijenhuis coordinates are just a
particular realization of this representation: the eigenvalues of $N$ are the
first $n$ coordinates and the remaining ones are constructed by quadratures
\cite{TMagri}. \par A vector field X is said to be bi-Hamiltonian w.r.t. $(P_0,
P_1)$ if there exist two smooth functions $h_0$ and $h_1$ such that 

\begin{equation} \label{eq:BHX} X=P_0\, dh_1=P_1\, dh_0 \ ,
\end{equation}
$d$ denoting the exterior derivative.
\par Under the above assumption on the eigenvalues of $N$, a BH vector field is
completely integrable, a set of independent involutive integrals being just the
eigenvalues $(\lambda_1, \ldots, \lambda_n)$ \cite{MM}. However, it has been
proved in \cite{B,F} that a very strong condition has to be satisfied by a
completely integrable system (Hamiltonian w.r.t. $P_0$) in order to have a BH
formulation in a neighborhood of an invariant Liouville torus, at least if one
searches for a second Poisson tensor compatible with $P_0$. Nevertheless, the
property of Liouville integrability, can be related with geometrical structures
which are actually different from the canonical BH structure. This can be done
in (at least) three distinct ways.

\begin{itemize}
\item[i)] Searching for a BH formulation without the canonical Poisson
structure
$P_0$ \cite{Mm1} (hereafter referred to as a {\it non canonical} BH
formulation). The BH structure constructed in Sect. 3 and the one considered in
Sect. 4 are just applications of such a construction. \item[ii)] Admitting a
degenerate BH formulation \cite{GZ}; for instance, this is the case of the rigid
body with a fixed point \cite{Ratiu,MP,Ian} and of the stationary flows of the
KdV hierarchy \cite{AFW,GT}. \item[iii)] Searching for a
quasi--bi--Hamiltonian formulation of $X$ \cite{francesi2,MT}. \end{itemize}

In connection with the third approach, we recall that the vector field X is said
to be quasi-bi-Hamiltonian w.r.t. $(P_0, P_1)$ \cite{francesi2} if there exist
three smooth functions $H$, $K$ and $\rho$ such that 

\begin{equation} \label{eq:QBHX} X=P_0\, dH=\frac{1}{\rho} P_1\, dK \ ;
\end{equation} in particular, if $\rho$ is the product of the eigenvalues of the
Nijenhuis tensor $N=P_1\, P_0^{-1}$, i.e., 

\begin{equation} \label{eq:Pfaff}
\rho=\prod_{i=1}^n \lambda_i \ ,
\end{equation} the QBH vector field X is said to be Pfaffian. \par It has beeen
proved in
\cite{francesi2} that any completely integrable system with two degrees of
freedom has a QBH formulation in a neighborhood of a Liouville torus.
Furthermore, for a Pfaffian QBH vector field with $n$ degrees of freedom we
proved \cite{MT} that the general solution of Eq.(\ref{eq:QBHX}), written in the
Darboux-Nijenhuis coordinates, is 

\begin{equation} \label{eq:HFnQBHS} H=\sum_{i=1}^n
\frac{f_i(\lambda_i;\mu_i)}{\prod_{j\neq i}(\lambda_i-\lambda_j)} \ , \qquad
K=\sum_{i=1}^n \frac{\rho_i\, f_i(\lambda_i;\mu_i)
}{\prod_{j\neq i}(\lambda_i-\lambda_j)} 
\quad\qquad (\rho_i:=\prod_{j\neq i}\lambda_j )
\end{equation} where each function$f_i$ is an arbitrary smooth function, at
most depending on one pair $(\lambda_i;\mu_i)$. A remarkable feature of $H$
and
$K$ is that they are separable (in the sense of Hamilton-Jacobi) as they verify
the Levi--Civita condition \cite{LC}, therefore the corresponding Hamilton
equations are integrable by quadratures. We stress the fact that, owing to the
arbitrariness of $f_i$, the functions $H$ and $K$ (\ref {eq:HFnQBHS}) provide a
class of separable functions different from the known St\"{a}ckel class (e.g.,
see \cite[p. 101]{P}). \par The previous results have been completed in
\cite{Bl1}, where it has been shown that a QBH vector field $X$ admits $n$
integrals of motion in involution $F_k \, (k=1,\ldots,n)$

\begin{equation} \label{eq:BlF} F_k= \sum_{i=1}^n \frac{\partial
c_k}{\partial\lambda_i}
\frac{f_i(\lambda_i;\mu_i)}{\prod_{j\neq i}(\lambda_i-\lambda_j)} \ ,
\end{equation} where $c_1,\ldots,c_n$ are the the coefficients of the minimal
polynomial of the Nijenhuis tensor $N$.
\begin{equation} \label{eq:Nmp}
\lambda^n +\sum_{i=1}^{n}c_{i}\lambda^{n-i}=\prod_{i=1}^n
(\lambda-\lambda_i) \ ; \end{equation} in particular, $F_1=-H, F_n=(-1)^{n}K$.
Furthermore, each function $F_k$ turns out to be separable. \par A natural
question arises about the mutual relations between the different formulations
of the above items. In this regard, we observe that a few examples of QBH
systems, such those considered in \cite{MT,Bl1} can be obtained as highly not
trivial reductions of degenerate BH systems \cite{GT}. However the relation
between the BH and the QBH formulation for a given vector field and the very
existence of one or both structures is not yet completely clarified. An open
problem, which we are not going to examine here, is to give conditions assuring
that a given integrable vector field with $n$--degrees of freedom admits
globally a QBH formulation. Since a theoretical result for $n> 2$ is still
lacking, it seems to us of some interest to collect and classify examples of
such systems. In this Letter our aim is just to discuss in some details an
explicit example of a system admitting both formulations. Its phase space is an
open dense submanifold of $R^{2n}$, so that its QBH and BH formulations are
globally defined. 

\section{the dynamical system and its canonical qbh formulation} 

Let $M=R^{2n}$, $ M \ni u=({\boldsymbol\lambda};{\boldsymbol\mu)}$. Let us
consider the Hamiltonian dynamical system $\dot u=X(u)$, with $X=P_0\, dH$;
$P_0$ is the canonical Poisson tensor and $H$ is given by

\begin{equation} \label{eq:CH} H=\sum_{i=1}^n
\frac{g_i(\lambda_i)}{\prod_{j\neq i}(\lambda_i-\lambda_j)} e^{a\mu_i} \ ,
\end{equation} where $g_i$ are arbitrary smooth functions, each one depending
only on the corresponding coordinate
$\lambda_i$, and
$a$ is an arbitrary constant.
\par The related Newton equations of motion take the form 

\begin{equation} \label{eq:Ne}
\ddot \lambda_k=2 \sum_{i\neq k}
\frac{\dot \lambda_i
\dot\lambda_k}{\lambda_k-\lambda_i}\qquad\qquad\qquad (k=1,\ldots,n)\ .
\end{equation} They were found to be solvable by F. Calogero
\cite{C2} and recently they have been shown \cite{C1} to describe a solvable
n-body systems in the plane. Indeed, as remarked in \cite{C1}, the previous
equations describe also a special case of the integrable relativistic
$n$--body problems introduced by S.N. Ruijsenaars and H. Schneider \cite{RS}.
\par Now, comparing (\ref{eq:CH}) with (\ref{eq:HFnQBHS}) one immediately
concludes that $\{{\boldsymbol\lambda};{\boldsymbol\mu}\}$ is a
Darboux--Nijenhuis chart for the Hamiltonian $H$, which is consequently
separable (a property unnoticed in \cite{C1}): 

\begin{prop} \label{pr:CQBH} The vector field $X=P_0\, dH$ is a Pfaffian QBH
vector field; $P_0$, $P_1$ and $N$ are given by (\ref{eq:PND}), $H$ is the
Hamiltonian (\ref{eq:CH}) and $K$ is given by 

\begin{equation} \label{eq:CK} K=\sum_{i=1}^n \frac{\rho_i\, g_i(\lambda_i)}
{\prod_{j\neq i}(\lambda_i-\lambda_j)} e^{a\mu_i} \ . \end{equation}
Furthermore, the corresponding Hamilton--Jacobi equation is separable; a
complete integral is
$S({\boldsymbol \lambda}; b_1,\ldots ,b_n)=-b_1t+W({\boldsymbol
\lambda};b_1,\ldots ,b_n)$ with

\begin{equation} \label{eq:W} W=\frac{1}{a} \sum_{i=1}^n \int^{\lambda_i}
\log\left(\frac{1}{g_i(\xi)}\sum_{j=1}^n b_j \, \xi^{n-j}\right) d\xi \ ,
\end{equation} and the Hamilton equations of motion can be solved by
quadratures. \qed
\end{prop} As it was shown in \cite{C2}, the Newton equations (\ref{eq:Ne}) can
be linearized by introducing a suitable set of coordinates. As a matter of fact,
these coordinates are strictly related with the Nijenhuis tensor (\ref{eq:PND}).
Indeed, let us consider the minimal polynomial of $N$ given by (\ref{eq:Nmp}):
expressing its coefficients ${\boldsymbol c}:=(c_1, \ldots, c_n)$ in terms of its
roots ${\boldsymbol \lambda}:=(\lambda_1, \ldots, \lambda_n)$ by means of the
Vi\`ete's formulae 

\begin{equation} \label{eq:Vieta}
\begin{split} c_1&=-\sum_{i=1}^n \lambda_i \\ c_2&=\sum_{i < j}^n \lambda_i
\lambda_j \\ c_3&=-\sum_{i < j < k}^n \lambda_i \lambda_j \lambda_k \\
.&\qquad . \\ .& \qquad . \\ c_n&= (-1)^n\prod_i^n \lambda_i \ ,
\end{split}
\end{equation} we can introduce the map

\begin{equation} \label{eq:Phi}
\Phi: {\boldsymbol \lambda}\mapsto {\boldsymbol c}:
c_k=\Phi_k(\lambda_1,\ldots,\lambda_n)\qquad (k=1,\ldots, n) \end{equation}
Now, we can easily obtain the following result: 

\begin{prop} \label{pr:cdot} The evolution of ${\boldsymbol c}$ along the flow
of $X$ is given by $\dot c_k=a F_k$, where
$F_k$ are given by (\ref{eq:BlF}), with $f_i(\lambda_i;\mu_i)=g_i(\lambda_i)\,
e^{a\mu_i}$. So, one has
$\ddot c_k=0$.

\begin{pf} On account of Eq. (\ref{eq:CH}), one gets 

\begin{equation} \label{eq:cdot}
\dot c_k=\sum_{i=1}^n \frac{\partial c_k}{\partial\lambda_i} \dot\lambda_i=
\sum_{i=1}^n \frac{\partial c_k}{\partial\lambda_i} \frac{\partial
H}{\partial\mu_i}= a\sum_{i=1}^n
\frac{\partial c_k}{\partial\lambda_i}
\frac{g_i(\lambda_i)}{\prod_{j\neq i}(\lambda_i-\lambda_j)}e^{a\mu_i}=a F_k
\ . \end{equation}
\end{pf}
\end{prop} This shows that the dynamics associated with $X$ is trivial when
expressed in terms of $ {\boldsymbol c}$, as it was already remarked in
\cite{C1,C2}. It seems to us of some interest to point out the algebraic meaning
of the map $\Phi$ (\ref{eq:Phi}), i.e., its relation with the Nijenhuis tensor
(\ref{eq:PND}).

\section{the dynamical system and its non canonical bh formulation} 

Let $\Psi:R^{2n}\rightarrow R^{2n}$ be the non canonical map 

\begin{equation}
\Psi: u=({\boldsymbol \lambda}; {\boldsymbol \mu}) \mapsto v=({\boldsymbol c}
;{\boldsymbol \gamma}) \qquad c_k=\Phi_k({\boldsymbol \lambda)} \ ;
\gamma_k=a F_k({\boldsymbol \lambda};{\boldsymbol \mu}) \ .
\end{equation} On account of Prop. \ref{pr:cdot} and of the fact that $F_k$ are
integrals of motion for $X$, it is
$\dot c_k=a F_k=\gamma_k$ and $\dot \gamma_k=a \dot F_k=0$; so, the vector
field
$X$ is mapped by $\Psi$ into the vector field $Y=({\boldsymbol \gamma}, 0)^T$
(of course, the whole QBH structure
$(P_0,P_1)$ could as well be transformed). Easily enough, for any dynamical
system $\dot v=Y(v)$ of this form one has the following result: 

\begin{prop} \label{pr:lBH} The system $\dot v=Y(v)$, with 
$Y=({\boldsymbol \gamma}, 0)^T$, is BH w.r.t. $ Q_0=
\begin{bmatrix} 0& I\\\ - I&0
\end{bmatrix} and \quad Q_1 =
\begin{bmatrix} 0&\Gamma\\\ -\Gamma &0
\end{bmatrix}
$, where $\Gamma:=diag(\gamma_1, \ldots, \gamma_n)$. The BH chain is 

\begin{equation} \label{BHc} Q_0\, dh_{j+1}=Q_1\, dh_j \qquad (j=0, 1,\ldots)
\ , \end{equation} with $h_0=\log (det \Gamma )$ and $h_j=
\frac{1}{2j}Tr (Q_1\,Q_0^{-1})^j=
\frac{1}{j}\sum_{i=1}^n \gamma_i^j \quad (j= 1, 2,3,\ldots) $. \qed
\end{prop}

\begin{rem}
\par For the sake of completeness, we recall that $X$ admits also a Virasoro
algebra of graded conformal symmetries $\tau_j\ (j=-1,0,1,\ldots)$ with 
$\tau_j=(0,{\boldsymbol\gamma^{j+1}})^T$ (${\boldsymbol
\gamma^{j}}:=(\gamma_1^j,\ldots,\gamma_n^j)^T$). The relation between this
algebraic structure and the BH structure is well known \cite{Oe}. $\qed$
\end{rem}
\par
\vspace{0.4cm} Since the BH structure can be transformed from the chart $\{
{\boldsymbol c}; {\boldsymbol \gamma}\}$ to the chart 
$\{ {\boldsymbol \lambda}; {\boldsymbol \mu}\}$, we conclude that the system
$\dot u=X(u)$, which is QBH w.r.t. $(P_0, P_1)$, is also BH w.r.t. $(Q_0,Q_1)$.
The Poisson tensors of the BH formulation can be obtained by transformation
(under $\Psi$) of the Poisson tensors $Q_0$,
$Q_1$ of Prop. \ref{pr:lBH}. We observe that the map $\Psi$ is non canonical
w.r.t. $Q_0$, so in the chart $\{ {\boldsymbol \lambda}; {\boldsymbol \mu}\}$ it
is $Q_0 \neq P_0$; as a matter of fact, the Poisson pair ($Q_0$, $Q_1$) take a
quite complicated form in the chart 
$\{ {\boldsymbol \lambda}; {\boldsymbol \mu}\}$, being simpler in the chart
$\{ {\boldsymbol c}; {\boldsymbol \gamma}\}$ (the opposite situation occurs
for the pair $(P_0,P_1)$).
\par For the sake of clarity, let us consider explicitly the case $n=2$. The
dynamical system $\dot u=X(u)$ is of the form (\ref{eq:QBHX}), with $H$ and
$K$  given respectively by (\ref{eq:CH}) and (\ref{eq:CK}): 

\begin{equation} H=\frac{g_1}{\lambda_{12}}
e^{a\mu1}-\frac{g_2}{\lambda_{12}}e^{a\mu_2} \ , \quad 
K=\frac{\lambda_2 g_1}{\lambda_{12}} e^{a\mu1}-\frac{\lambda_1
g_2}{\lambda_{12}}e^{a\mu_2}
\quad (\lambda_{12}:=\lambda_1-\lambda_2)\ .
\end{equation} So, we have

\begin{equation} \label{eq:CS2D}
\begin{split} &\dot \lambda_1=\frac{a g_1}{\lambda_{12}} e^{a\mu_1} \\
&\dot \lambda_2=-\frac{a g_2}{\lambda_{12}}e^{a\mu_2} \\ &\dot
\mu_1=-\frac{\partial}{\partial\lambda_1}(\frac{g_1}{\lambda_{12}}
)e^{a\mu_1} -\frac{g_2}{\lambda_{12}^2} e^{a\mu_2} \\ &\dot
\mu_2=-\frac{g_1}{\lambda_{12}^2} e^{a\mu_1}
-\frac{\partial}{\partial\lambda_2}(\frac{g_2}{\lambda_{12}}) e^{a\mu_2}\ .
\end{split}
\end{equation} This system is separable; a complete integral of the
corresponding Hamilton-Jacobi equation is
$S(\lambda_1,\lambda_2;b_1,b_2)=-b_1 t+W$. According to (\ref{eq:W}), $W$
is given by

\begin{equation} W=\frac{1}{a}\int^{\lambda_1} d\xi \,
\log\frac{b_2+b_1\xi}{g_1(\xi)}+ \frac{1}{a}\int^{\lambda_2} d\xi \,
\log\frac{b_2+b_1\xi}{g_2(\xi)} \ , \end{equation} therefore the general
solution of Hamilton equations is 

\begin{equation}
\begin{split} t-t_0&=\frac{1}{a}\int^{\lambda_1} d\xi \,
\frac{\xi}{b_2+b_1\xi}+ \frac{1}{a}\int^{\lambda_2} d\xi \,
\frac{\xi}{b_2+b_1\xi} \\ \beta&=\frac{1}{a}\int^{\lambda_1} d\xi \,
\frac{1}{b_2+b_1\xi}+ \frac{1}{a}\int^{\lambda_2} d\xi \,
\frac{1}{b_2+b_1\xi} \\ \mu_1&=\frac{1}{a}
\log\frac{b_2+b_1\lambda_1}{g_1(\lambda_1)} \\
\mu_2&=\frac{1}{a}\log\frac{b_2+b_1\lambda_2}{g_2(\lambda_2)} \ .
\end{split}
\end{equation}

As for the map $\Psi$, it is

\begin{equation} \label{eq:psimap} c_1=-(\lambda_1+\lambda_2) \ , \quad
c_2=\lambda_1\lambda_2 \ , \quad
\gamma_1=-aH \ , \quad
\gamma_2=aK \ ;
\end{equation} the BH dynamical system $\dot v=Y(v)$ is given by 

\begin{equation}
\dot c_1=\gamma_1 \ , \quad
\dot c_2=\gamma_2 \ , \quad
\dot \gamma_1=0 \ , \quad
\dot \gamma_2=0 \ ,
\end{equation} with Hamiltonians
\begin{equation} h_1=\gamma_1+\gamma_2\ ,\qquad
h_2=\frac{1}{2}(\gamma_1^2+\gamma_2^2) \ . \end{equation} In order to
transform this BH structure, it suffices to consider the inverse map $\Psi^{-1}$
given by

\begin{equation} \label{eq:Nmap3D}
\begin{split}
\lambda_{1,2}&=-\frac{1}{2}c_1\pm\frac{1}{2}(c_1^2-4c_2)^{1/2} \ , \\
\mu_{1,2}&=\frac{1}{a}
\log (c_1\gamma_1-2\gamma_2\mp\gamma_1
(c_1^2-4c_2)^{1/2})-\frac{1}{a}\log(2a) \ . \end{split}
\end{equation} We obtain $h_1=a(K-H)$, $h_2=\frac{1}{2}a^2(K^2+H^2)$ and 

\begin{equation} \label{eq:QC2D} Q_0 =
\begin{bmatrix} 0&A\\\ -A^T&B
\end{bmatrix} \ , \quad Q_1 =
\begin{bmatrix} 0&C\\\ -C^T&D
\end{bmatrix} \ ;
\end{equation}
$A$ and $C$ are $(2 \times 2)$ matrices with entries 

\begin{equation}
\begin{split}
&A_{ij}=\frac{(-1)^{i+1}}{a^2\lambda_{12}}\frac{1+\lambda_i\lambda_j}
{g_j}e^{-a\mu_j}
\qquad (i,j=1,2) \\ & C_{ij}=(-1)^{i+j}\left(\lambda_{j+1}
-\lambda_i\lambda_j+\lambda_j(\lambda_i-1) \frac{g_{j+1}}{g_j}
e^{a(\mu_{j+1}-\mu_j)} \right) \quad (i,j=1,2) \\ \end{split}
\end{equation} where $\lambda_3:=\lambda_1, g_3:=g_1,\mu_3:=\mu_1$; 
$B$ and $D$ are
$(2 \times 2)$ skew-symmetric matrices with 

\begin{equation}
\begin{split} B_{21}&=
\frac{(1+\lambda_1\lambda_2)\lambda_{12}}{a^3g_1g_2}
\left(\frac{\partial}{\partial\lambda_1}(\frac{g_1}{\lambda_{12}^2}
)e^{-a\mu_2}+ \frac{\partial}{\partial\lambda_2}(\frac{g_2}{\lambda_{12}^2}
)e^{-a\mu_1}\right )\\  D_{21}&= \frac{1}{a^2 \lambda_{12}^3g_1 g_2 } \left(
(2g_2+g_2'
\lambda_{12})\left(g_1\lambda_2(1-\lambda_1)+g_2\lambda_1(\lambda_2-1)
e^{a(\mu_2-\mu_1)}\right ) \right)\\ &-\frac{1}{a^2 \lambda_{12}^3g_1 g_2}
\left
((2g_1-g_1'\lambda_{12})\left(g_2\lambda_1(\lambda_2-1)+g_1\lambda_2
(1-\lambda_1) e^{a(\mu_1-\mu_2)} \right)
\right) \ .
\end{split}
\end{equation} Clearly, this BH structure is much more involved than the QBH
structure $(P_0,P_1)$.

\section{Concluding remarks} The dynamical system considered by F. Calogero
in \cite{C1}, as well as its generalization analyzed in this Letter, is naturally a
Pfaffian QBH system; consequently, it is separable and therefore integrable by
quadratures, the physical coordinates being just the Darboux-Nijenhuis
coordinates w.r.t. $(P_0, P_1)$ (as for other examples of systems which admit
this type of formulation, see \cite{FF,Bl2}). \par A property of this system,
which seems to be rather peculiar, is that the minimal polynomial of the
Nijenhuis tensor $N$ directly suggests the introduction of a non canonical map
which provides the non canonical BH structure $(Q_0,Q_1)$; this procedure is an
example of the strategy mentioned in item i) of the Introduction. In this regard,
it is known that any Liouville integrable system can be given infinitely many
non canonical BH formulations \cite{Mm1} in a neighborhood of an invariant
Liouville torus; to this end, a non canonical map has to be constructed starting
from a set of action-angles variables \cite{Mm2}. This procedure is quite
similar to the one we have used in this Letter; the main difference is given by
the fact that we can construct the non canonical map between the
Darboux--Nijenhuis coordinates $\{{\boldsymbol\lambda}; {\boldsymbol
\mu}\}$ and the coordinates $\{ {\boldsymbol c}; {\boldsymbol\gamma}\}$
directly, without passing through the action--angle variables. \par As a final
remark, let us observe that a quite similar situation occurs if one considers
the dynamical system 

\begin{equation} \label{Bc}
\dot u=P_0\, dH \ ,\qquad u^T=(s_1, s_2,\theta_1, \theta_2) \ ,\qquad
H=s_1(1+s_2^2) \ , \end{equation} where $P_0$ is the canonical Poisson tensor,
$(s_1,s_2)$, $(\theta_1, \theta_2)$ are action-angles variables respectively.
This system was introduced by Brouzet
\cite{B} as a counterexample to the existence of a canonical BH structure.
However, since $n=2$, it can be endowed with a canonical (not Pfaffian) QBH
structure, which is explicitly given by \begin{equation}
\dot u=\frac{1}{H}\, P_1\, dK \ ,\qquad K=-\frac{1}{s_1}+2 \arctan s_2 \
,\qquad P_1=
\begin{bmatrix} 0& S\\\ - S&0
\end{bmatrix} \ ,
\end{equation} with $S=diag(s_1,s_2)$. Thus, the Brouzet's counterexample has
both a non canonical BH structure and a canonical QBH one, as well as the
system we have considered in this Letter.

\vspace{0.9truecm} {\bf Acknowledgments.}

We gratefully acknowledge many valuable criticisms by an anonymous referee.
One of us (G.T.) takes the occasion to thank F.Calogero and J.-P. Francoise for
enlightening discussions. \par This work has been partially supported by the
GNFM of the Italian CNR and by the project "Metodi geometrici e probabilistici
in Fisica Matematica" of the Italian MURST.

\end{document}